\RequirePackage[]{silence}
\documentclass[11pt]{article}

\usepackage[letterpaper, margin=1.00in]{geometry}
\usepackage{amsmath, amssymb, amsthm}
\usepackage{cite}
\usepackage{url}
\usepackage{appendix}
\usepackage{graphicx}
\usepackage{color}
\usepackage[plain]{algorithm}
\usepackage{enumitem}
\usepackage[noend]{algpseudocode}
\usepackage{float}
\usepackage{todonotes}
\usepackage[framemethod=tikz]{mdframed}
\usepackage[capitalize]{cleveref}

\newtheorem{theorem}{Theorem}[section]
\newtheorem{lemma}[theorem]{Lemma}

\newtheorem{remark}[theorem]{Remark}

\newtheorem{proposition}[theorem]{Proposition}
\newtheorem{definition}[theorem]{Definition}

\Crefname{thm}{Theorem}{Theorems}
\Crefname{lemma}{Lemma}{Lemmas}
\Crefname{claim}{Claim}{Claims}
\Crefname{remark}{Remark}{Remarks}

\newtheorem*{theoremRinner}{Theorem~\ref{\repthmref}}
\newenvironment{theoremR}[1]
  {\def\repthmref{#1}\theoremRinner (restated)}{\endtheoremRinner}

\renewcommand{\paragraph}[1]{\vspace{0.15cm}\noindent {\bf #1}}

\algnewcommand\algorithmicswitch{\textbf{switch}}
\algnewcommand\algorithmiccase{\textbf{case}}

\algdef{SE}[SWITCH]{Switch}{EndSwitch}[1]{\algorithmicswitch\ #1\ \algorithmicdo}{\algorithmicend\ \algorithmicswitch}%
\algdef{SE}[CASE]{Case}{EndCase}[1]{\algorithmiccase\ #1}{\algorithmicend\ \algorithmiccase}%
\algtext*{EndSwitch}%
\algtext*{EndCase}%

\algnewcommand\algorithmicwithprob{\textbf{with probability}}
\algnewcommand\algorithmicotherwise{\textbf{otherwise}}

\algdef{SE}[WithProb]{WithProb}{EndWithProb}[1]{\algorithmicwithprob\ #1\ \algorithmicdo}{\algorithmicend\ \algorithmicwithprob}%
\algdef{Ce}[Otherwise]{WithProb}{Otherwise}{EndWithProb}
  {\algorithmicotherwise}%
\algtext*{EndWithProb}%
\algtext*{EndOtherwise}%

\newcommand{\eps}{\epsilon}

\newcommand{\floor}[1]{\lfloor #1 \rfloor}
\newcommand{\ceil}[1]{\lceil #1 \rceil}

\newcommand{\FullOrShort}{full}

\ifthenelse{\equal{\FullOrShort}{full}}{
	
	  \newcommand{\fullOnly}[1]{#1}
		
	  \newcommand{\shortOnly}[1]{}

  }{
		\usepackage{times}
    
	  \newcommand{\fullOnly}[1]{}
	  \newcommand{\shortOnly}[1]{#1}
  }

\begin{document}

\date{}

\author{Noga Alon\thanks{\texttt{nogaa@post.tau.ac.il}}\\ Tel Aviv University \and Mohsen Ghaffari\thanks{\texttt{ghaffari@mit.edu}}\\ MIT \and Bernhard Haeupler\thanks{\texttt{haeupler@cs.cmu.edu}}\\ CMU \and Majid Khabbazian \thanks{\texttt{mkhabbazian@ualberta.ca}}\\ University of Alberta}
\title{Broadcast Throughput in Radio Networks: \\ Routing vs. Network Coding}
\maketitle

\setcounter{page}{0}
\thispagestyle{empty}
\begin{abstract}
The \emph{broadcast throughput} in a network is defined as the average number of messages that can be transmitted per unit time from a given source to all other nodes when time goes to infinity. 

Classical broadcast algorithms treat messages as atomic tokens and \emph{route} them from the source to the receivers by making intermediate nodes store and forward messages. The more recent \emph{network coding} approach, in contrast, prompts intermediate nodes to mix and code together messages. It has been shown that certain wired networks have an asymptotic \emph{network coding gap}, that is, they have asymptotically higher broadcast throughput when using network coding compared to routing. Whether such a gap exists for wireless networks has been an open question of great interest. We approach this question by studying the broadcast throughput of the \emph{radio network model} which has been a standard mathematical model to study wireless communication. 


We show that there is a family of radio networks with a tight $\Theta(\log \log n)$ network coding gap, that is, networks in which the asymptotic throughput achievable via routing messages is a $\Theta(\log \log n)$ factor smaller than that of the optimal network coding algorithm. We also provide new tight upper and lower bounds that show that the asymptotic worst-case broadcast throughput over all networks with $n$ nodes is $\Theta(1 / \log n)$ messages-per-round for both routing and network coding. 


\end{abstract}
\newpage 

%
%
%
\section{Introduction}

\emph{Broadcasting}, that is, transmitting one or multiple messages from a source to some or all nodes in a network, is one of the most important network communication primitives. It is particularly interesting in multi-hop wireless networks: While wireless communication is of broadcast-type on a local level (transmissions reach all close-by nodes), collision and interference make global broadcasts over multiple hops challenging. 

The \emph{radio network} model~\cite{CK} was designed to capture these characteristics and has become one of the standard mathematical models to study wireless network communication. In this model, communication occurs in synchronous \emph{rounds} in which each node in a network, represented by a graph $G$, can decide to \emph{send a packet} or \emph{listen}. Nodes that listen receive the packet of a sending neighbor if there is exactly one sending neighbor. On the other hand, if two or more neighbors of a node $v$ send simultaneously, then their transmissions \emph{collide}, that is, interfere and are useless for $v$. 

In this paper we study what \emph{broadcast throughput} is possible in radio networks. That is, we want to know how many messages per round can a source broadcast to all nodes on average as time (and the number of messages) goes to infinity. In addition to determining the optimal broadcast throughput we also want to know whether any coding is necessary to achieve it. 

The reason for differentiating coding and non-coding approaches is due to the following historical developments: The classical way of transmitting multiple messages over a networks is \emph{routing}, that is, messages are regarded as atomic tokens and routed from the source to the receivers by making intermediate nodes store and forward messages. It was long known that routing can achieve the optimal throughput for point-to-point communications. However it was not until 2000 that Ahlswede et al. discovered that the broadcast throughput of a network can be increased by having intermediate nodes code messages together. This \emph{network coding} approach has since let to both fundamental new insights and simple, distributed and throughput-optimal broadcast algorithms for many settings. Wireless networks, in particular, have become a popular setting to study the impact of network coding and both positive and negative results have been reported.

The situation is similarly open for broadcast in the radio network model. Classical broadcast algorithms achieve a $\Theta(1/\log^2 n)$ messages-per-round throughput in any $n$-node network. Recently algorithms with $\Theta(1/\log n)$ messages-per-round throughput have emerged, all of which employ network coding. See \Cref{subsec:related} for related work. This prompts several questions: How much (if any) advantage can network coding provide over routing in radio networks? What is the optimal worst-case broadcast throughput in radio networks? Is network coding necessary to achieve it?

We address these questions and show that there is a family of radio networks with a tight $\Theta(\log \log n)$ network coding gap, that is, networks in which the asymptotic throughput achievable via routing messages is a $\Theta(\log \log n)$ factor smaller than that of the optimal network coding algorithm. Surprisingly, we also provide new tight upper and lower bounds that show that the asymptotic worst-case broadcast throughput over all networks with $n$ nodes is $\Theta(1 / \log n)$ messages-per-round for both routing and network coding.

\subsection{Related Work}\label{subsec:related}
We present a brief review of closely related works, divided into three parts: \emph{broadcast problem in radio networks}, study of \emph{network coding advantage}, and finally  \emph{network coding in wireless networks}.

\paragraph{Broadcast in Radio Networks:} The study of broadcast in radio networks has a long line of history, dating back to 1985 work of Chalamatac and Kutten\cite{CK}. 
As a result of about 20 years of research, worst-case optimal single-message broadcast time complexity is well-understood:  $\Theta(D\log \frac{n}{D}+\log^2 n)$ for unknown-topology \cite{KP, CR, ABLP, KM93,BGI1} and $\Theta(D+\log^2 n)$ for known topology~\cite{KP-DC, GPX, ABLP}, where $D$ is the network diameter.
Peleg~\cite{Peleg07} provides a nice survey.  

For multiple-message broadcast, a summary with a focus on throughput is as follows\footnote{We remark that, quite a few papers study multi-message broadcast without considering any packet size bounds. This is completely irrelevant to the case in this paper and thus, we do not mention them here.}: Bar-Yehuda et al.~\cite{BII93} used the \emph{Decay broadcast protocol} of \cite{BGI1} to get a $k$-message broadcast algorithm with dependency of $O(k \log^2 n)$ rounds on the number of messages $k$, i.e., throughput of $\Omega(1/\log^2 n)$ messages-per-round. This routing-based throughput remained the best known for about two decades, recurring in many papers\footnote{Some of these papers are about the also-widely-studied \emph{gossiping problem}, a.k.a., \emph{all-to-all broadcast}, which from the worst-case throughput point of view, is equivalent to an $n$-message broadcast problem.}, until recently where network coding was shown to achieve dependency $O(k\log n)$~\cite{KK, GHK13}, i.e., throughput of $\Omega(1/\log n)$ messages-per-round. A routing-based $\Omega(1/\log n)$ throughput was claimed in \cite{FQ06} but its correctness was disproved \cite{QinPersonalCommunication}.

\paragraph{Network Coding Advantage:} Since its introduction in \cite{ACLY} network coding has become a well-studied subfield of information theory. Most related are studies that study the \emph{network coding gap} (i.e., the ratio of the optimal throughput using network coding to that of routing) for different network models (see, e.g.,~\cite{Li-Li-Lau, agarwal2004advantage, Langberg, Goel}). This network coding gaps are often deeply connected to combinatorial or graph theoretical problems. In wired \emph{undirected} networks, using a classical edge-disjoint spanning trees result of Tutte and Nash-Williams, Li et al.~\cite{Li-Li-Lau} show this gap to be at most a constant of $2$. On the other hand, for \emph{directed} wired networks,
Agarwal and Charikar \cite{agarwal2004advantage} show the coding gap to exactly correspond to the integrality gap of the (directed) Steiner-tree LP and use an integrality gap result of Halperin et al. \cite{halperin2003integrality} to prove an $\Omega((\log n / \log \log n)^2)$ bound on the coding gap for \emph{directed} networks. Whether the gap for directed wired networks is polylogarithmic or even polynomial in $n$ is a major open question. Recently, Censor-Hillel et al.\cite{CHGK13} show a tight gap of $\Theta(\log n)$ for the model where in each round, each node can send one packet to all of its neighbors (no collisions).

\paragraph{Network Coding in Wireless Networks:} Whether network coding offers advantages in wireless networks has become a question of both practical and theoretical interest. A prominent example is the work of Katti et al.~\cite{Katti06} which implemented a network coding strategy for practical wireless networks and reported significant constant factor throughput improvements. Following  \cite{Katti06} many papers have studied different aspects of network coding in wireless networks, such as, energy-efficiency, robustness to packet losses, dynamic networks, etc. (e.g. \cite{NTNB, FWBY, Goel}).

\section{Setup}\label{sec:setup}
\subsection{Model}
We consider the well-studied \emph{radio network model}, first introduced by Chlamtac and Kutten~\cite{CK}. The connections in this model are presented by a graph $G=(V,E)$ with $|V|=n$. The communication occurs in synchronous rounds where in each round, each node either transmits a packet with length $B=\Theta(\log n)$ bits or listens. Each node receives a packet if and only if it is listening and exactly one of its neighbors is transmitting a packet. In particular, if two or more neighbors of a node $v$ transmit simultaneously, their transmissions interfere (collide) at $v$ and $v$ does not receive anything.

\subsection{Problem Statement}\label{subsec:problem}

\paragraph{Broadcast Problem:} We study the \emph{$k$-message broadcast} problem in which one source node broadcasts $k$ messages to all other nodes. Formally an instance of the \emph{$k$-message broadcast} problem consists of a graph $G$, a source node $s$ and $k$ \emph{messages} consisting of $B$ bits each. The messages are initially only known to the source $s$ and the goal is to deliver all messages to all nodes of $G$ in as few rounds as possible. 

We emphasize that, we use the term \emph{message} to indicate the blocks of information related to the problem that originally reside in the source node, and we use the term \emph{packet} to indicate the blocks of information that are transmitted by nodes throughout the algorithm. The difference becomes more clear as we next define routing-based algorithms and network coding algorithms.

\paragraph{Routing-Based Algorithms:} In a routing-based algorithm, messages are viewed as (atomic) tokens and nodes can only store and forward them. Each packet contains therefore exactly one message. It is easiest to think of routing algorithms in the following way: In each round $r$, each node $v$ has a buffer which contains all the messages that $v$ has received by the end of round $r-1$. Initially, these buffers are empty except for source's buffer which contains all $k$ messages. In each round $r$, for each node $v$, if $v$ transmits a packet $p$, then packet $p$ has to be equal to one of the messages that is in the buffer of node $v$ in that round. When a node $w$ receives a packet, the related message gets added to the buffer of $w$.

\paragraph{Network Coding Algorithms:} In contrast to routing based algorithms, in network coding algorithms, packets can be different from messages and each packet can contain (partial) information about many messages, as long as the total length of the packet is $B$ bits. More precisely, in each round $r$, each node $v$ has a buffer which contains all the packets that $v$ has received by the end of round $r-1$. Again, initially, these buffers are empty except for source node's buffer which contains all the $k$ messages (as $k$ packets). In each round $r$, for each node $v$, if $v$ transmits a packet $p$, then packet $p$ can be any function of all the packets that are in the buffer of $v$ in that round, subject to the condition that $p$ contains at most $B$ bits. Note that routing-based algorithms are a special case of network coding algorithms.

\paragraph{Broadcast Throughput:} For a given graph $G$ and a source node $s$, the routing-based broadcast throughput is defined\footnote{\label{note}It is easy to see that the limits in the definitions of throughput exist. See \Cref{prop:limit} in \Cref{app:setup}.} as $\lim_{k\rightarrow \infty}\frac{k}{T^R_k(G, s)}$, where $T^R_{k}(G,s)$ is the smallest number of rounds required for broadcasting $k$ messages from source $s$ to all nodes in $G$, when using routing-based algorithms. Similarly, the network coding broadcast throughput is defined as $\lim_{k\rightarrow \infty}\frac{k}{T^{NC}_k(G, s)}$, where $T^R_{k}(G,s)$ is the smallest number of rounds required for broadcasting $k$ messages from source $s$ to all nodes in $G$, when using network coding algorithms. The network coding gap for a graph $G$ and source node $s$ is defined as $\lim_{k\rightarrow \infty} \frac{T^R_k(G, s)}{T^{NC}_k(G, s)}$. 

\subsection{Broadcast on Bipartite Networks}\label{subsec:bipartite}
We next define broadcast problem on bipartite networks and argue that studying bipartite networks is enough for understanding the worst-case broadcast throughput in general graphs.

\paragraph{Bipartite Networks:}
In a bipartite network $H=(V, E)$, set of vertices $V$ is composed of two disjoint nonempty sets $A$ and $B$, which are each an independent set of $H$. We call the nodes in $A$ and $B$ respectively \emph{senders} and \emph{receivers}. We often use numbers $1$ to $\eta$ to identify senders and thus have $A=\{1, 2, \dots, \eta\}$. Note that $B$ corresponds to a set of subsets of $A$, one subset for each receiver $u$, representing simply neighbors of $u$. Without loss of generality, we can assume that each subset appears only once as if two receivers have the same sender neighbors, they always receive the same packets and thus, studying just one is enough for understanding the throughput problem.

\paragraph{Bipartite Broadcast Problem:} We define the $k$-message broadcast problem on bipartite networks as all senders initially have all the $k$ messages and the goal is to deliver all messages to all receivers. Bipartite Broadcast Throughput is defined similarly.

\medskip
Next, we argue that studying worst-case bipartite broadcast throughput is enough for understanding the worst-case broadcast throughput in general graphs (up to constant factors). 

\begin{theorem}\label{prop:goingToBipartite} For both routing and network coding, the worst-case optimal throughput on $n$-node bipartite networks is, up a constant factor, equal to the worst-case optimal throughput general graphs with $n+1$ nodes.
\end{theorem}

Here is a rough sketch of the argument: Each bipartite broadcast problem can be turned into an almost-equal general broadcast problem by just adding one source $s$ and connecting it to all senders $A$. The converse relation is by decomposing each graph into bipartite networks where, for each $d$, nodes at distances $d$ and $d+1$ from source define one bipartite network. By pipelining over these bipartite networks, a high-throughput bipartite broadcast algorithm gives a high throughput general broadcast algorithm. 

\fullOnly{
\begin{proof}
To prove the theorem, we show two things: 
\begin{itemize}
\item[(a)] If there is a bipartite network with $n$ nodes for which any algorithm has bipartite broadcast throughput at most $x$ messages-per-round, then there is a network with $n+1$ nodes for which any algorithm has throughput at most $x$ messages-per-round. 

\item[(b)] If for any bipartite network with $n$ nodes we have a broadcast algorithm with throughput of at least $x$ messages-per-round, then for any general network with $n+1$ nodes we have a broadcast algorithm with broadcast throughput at least $\Theta(x)$ messages-per-round.
\end{itemize}

For part (a), simply add a source node $s$---which initially contains all $k$ messages---to the bipartite network and connect it to all senders. If a broadcast algorithm delivers $k$ messages to all nodes in at most $\frac{k}{x}$ rounds, then repeating the same transmissions in the bipartite network part gives a bipartite broadcast algorithm with throughput at least $x$.

For part (b), note that for each general graph $G$, the broadcast problem on $G$ can be broken into broadcast problems on a set of bipartite networks. In particular, if we have a broadcast algorithm with throughput $y$ for all bipartite networks with at most $n$ nodes, then we can achieve a throughput of at least $\Theta(y)$ in general graphs with $n+1$ nodes. Consider an arbitrary graph $G$ with source node $s$ and the Breadth First Search (BFS) layering of $G$ where the $i^{th}$ layer contains all the nodes at distance $i$ from the the source $s$. Each two consecutive BFS-layers define a bipartite network. We divide the messages into batches of $k'=\frac{k}{D}$. Delivering each batch from one layer to the next takes $O(k'/y)$ rounds. By spacing the progress of batches $3$ layers apart, we can pipeline different batches simultaneously and deliver all batches to all nodes in $O((D + \frac{k}{k'}) \cdot \frac{k'}{y}) = O(\frac{k}{y})$ rounds, thus achieving a throughput of $\Theta(y)$. 
\end{proof}
}

%

Lastly we remark that when looking at throughput there is no difference between distributed and centralized algorithms since the time to learn the topology is independent from the number of messages $k$ to be broadcast. Similarly the existence of randomization, IDs and collision detection becomes irrelevant as well.

\section{Overview and Our Results}\label{subsec:results}
In this section, we give an overview of the structure of related known upper and lower bounds, which will be helpful for understanding our approach. After that, we state our results formally.

\subsection{Background}


When trying to transmit information in a bipartite network, there is a dichotomy between reaching receivers with small and large degrees. In particular, when many senders transmit, one would expect to reach receivers with small degrees while causing collisions for receivers with large degrees. On the other hand, fewer senders transmitting leads to fewer collisions for receivers with large degrees but might miss many small-degree receivers. 

As we see next, this dichotomy is very apparent in both upper and lower bounds. In particular, all known algorithms divide senders into $\Theta(\log n)$ \emph{degree-ranges}, where the degrees of receivers in a degree-range are equal, up to a constant factor. Similarly, impossibility results, such as \cite{ABLP}, show this to be essentially necessary:

\paragraph{Upper Bound for Broadcasting One Message:} An easy solution for transmitting one message in a bipartite network is the Decay protocol \cite{BGI1}, which works as follows:

\begin{mdframed}[hidealllines=true,backgroundcolor=gray!20]
\centering
\begin{minipage}{15cm}
\begin{algorithm}[H]
\begin{algorithmic}[1]
\small

\Statex

\For{$i=1$ to $\Theta(\log n)$}
	\For{$j=1$ to $\Theta(\log n)$}
		\WithProb{$\frac{1}{2^i}$} \ \textsc{Transmit} message $m$; \ \textbf{otherwise} \ remain \textsc{silent}
		\EndWithProb
	\EndFor
\EndFor
\label{alg:NCUB}
\end{algorithmic}
\end{algorithm} 
\end{minipage}
\end{mdframed}

\smallskip

It is easy to see that in each iteration of the inner loop, each receiver that has degree in $[2^{i-1}, 2^i]$ receives the message with constant probability. Repeating this $\Theta(\log n)$ times leads to each such receiver receiving the message with high probability and thus, each complete run of the inner-loop covers one receiver degree-range. The outer loop then covers all receiver degrees from $1$ to $n$.

\paragraph{Lower Bound for Broadcasting One Message:} The question is of course whether the $\Theta(\log^2 n)$ rounds of the Decay protocol can be improved. In a technically challenging lower bound, Alon et al.~\cite{ABLP} showed that delivering the message to all receivers of each fixed degree-range requires $\Theta(\log n)$ rounds in the worst case. Furthermore, their proof shows that transmissions aimed at different degree-ranges essentially do not help each other. This completes the $\Theta(\log^2 n)$ round lower bound. 

\medskip

\paragraph{Upper Bounds for Broadcasting Multiple Messages:} When trying to transmit $k > 1$ messages, there are two easy ways to extend the Decay-Protocol. 

\medskip

Firstly, one can repeat the Decay-Protocol for each of the $k$ messages. For example, repeating the inner-loop of the Decay for each of the $k$ messages delivers all messages to all receivers within degree-range in $\Theta(k \log n)$ rounds. Repeating this for each of the $\log n$ degree-ranges leads to a  $\Theta(k \log^2 n)$ round complexity or, differently phrased, to a throughput of $\Theta(1 / \log^2 n)$ messages-per-round (see \Cref{app:ncub}).

\medskip
 
Secondly, to speed this up, one could apply the inner-loop for each message only once instead of $\Theta(\log n)$ times. This results in each receiver receiving each message with constant probability and it is easy to see that for $k = \Omega(\log n)$ messages, each receiver will receive a constant fraction of all messages, with high probability. Unfortunately, each receiver might receive a different set of messages. Indeed, for many networks it is quite likely that no message is received by all receivers. Forward-error-correcting codes~\cite{macwilliams1977theory} offer an easy solution to this problem. For any $k$ messages one can, for example, create $100 k$ coded messages of equal length such that the original messages are decodable from any subset of $k$ coded messages. Therefore, using coding in the inner loop one can broadcast $k$ messages using $100 k$ rounds to all receivers of a degree-range. Again, iterating this over all $\log n$ degree-ranges leads to a total of $\Theta(k \log n)$ rounds for a throughput of $\Theta(1 / \log n)$ messages-per-round in any bipartite network. That is, a throughput of $\Theta(1)$ for each degree-range.

\subsection{Our Results}
The first (small) contribution this paper is to give a simple and direct proof showing that the above coding approach is optimal in a worst-case network:

\begin{theorem}\label{thm:NCLB} There exists a radio network where for any $k$, broadcasting $k$ messages requires at least $\Omega(k\log n)$ rounds. That is, the throughput for any broadcast algorithm is $O(1/\log n)$ messages-per-round.
\end{theorem}

The proof is very intuitive: For each degree-range, one can deliver at most $\Theta(1)$ messages-per-round and transmissions to each of the $\Theta(\log n)$ different degree-ranges do not help each other (much). 

With the throughput of network coding fully understood, the main question remaining is whether this $\Theta(1)$ messages-per-round throughput per degree-range can also be achieved with routing. Given the prior work, the two most plausible answers to this question, which both would be interesting, are as follows:

\smallskip

The first case is that, it is possible to assign uncoded messages to senders such that a constant fraction of messages is received by all receivers. This would lead to a $\Theta(1 / \log n)$ messages-per-round throughput for any (bipartite) network using only routing. This would show that routing is asymptotically as efficient as network coding.

The second plausible case is that, it is not possible to transmit $k$ messages to all receivers of a fixed degree-range in less than the $\Theta(k \log n)$ rounds if one only forwards uncoded messages. If this is true for each of the $\log n$ degree-ranges individually, then, following the argument line of \cite{ABLP} and \Cref{thm:NCLB}, one would expect that again transmissions for separate degrees can not help each other much. This would lead to an $O(1 / \log^2 n)$ messages-per-round bound on the throughput of routing and would show that network coding is strictly necessary for asymptotically optimal broadcast algorithms in radio networks. 

\medskip
We show neither of these to be the case. For networks with receivers of one degree-range, we show that routing can achieve a throughput of at most $O(1/\log\log n)$ messages-per-round. This stands in contrast to the $\Theta(1)$ messages-per-round throughput of network coding for such networks and proves an $\Omega(\log\log n)$ network coding gap in such networks:

\begin{theorem}\label{crl:gap}There exist bipartite networks with $n$ nodes where all receivers have degrees in range $[d/2, 2d]$, for $d>n^{0.01}$, and for which the broadcast throughput of any routing algorithm is at most $O(1/\log\log n)$ messages-per-round. For any such network, the network coding throughput is $\Theta(1)$ messages-per-round.
\end{theorem} 

It turns out that this $\Theta(\log \log n)$ gap is tight for bipartite graphs with one degree-range:
\begin{theorem}\label{lem:degschedule2}
For any bipartite network with $n$ nodes such that all receivers have degrees in range $[d/2, 2d]$, there exists a routing scheme with throughput $\Theta(1/\log \log n)$.
\end{theorem}

Lastly, we show that, surprisingly, this $\Theta(\log \log n)$ bound does \emph{not} add up for different degree-ranges and the gap disappears (asymptotically) for the worst-case over all $n$-node graphs:

\begin{theorem}\label{lem:globalschedule}
For any bipartite network $G$ with $n$ nodes there exists a routing scheme with throughput $\Omega(1 / \log n)$ messages-per-round.
\end{theorem}

\section{Network Coding}\label{sec:NC}
\shortOnly{In this section, we sketch the proof of \Cref{thm:NCLB}. Details appear in \Cref{app:NCLB}.}
\fullOnly{In this section, we present the proof of \Cref{thm:NCLB}.}

As discussed in \Cref{subsec:results}, our proof follows the spirit of the $\Omega(\log^2 n)$ lower bound of \cite{ABLP} for broadcasting a single message in a bipartite network. The intuition of the proof is that it is essentially unavoidable to aim at different degree ranges in different rounds. In particular, in a random bipartite network with different receiver degrees any round in which an $\alpha$ fraction of senders is sending will effectively only be useful for receivers of degree around $\frac{1}{\alpha}$. More precisely, the probability (and thus the number) of receptions is dropping exponentially as receiver degrees go away from $\frac{1}{\alpha}$: 

%


\shortOnly{
\begin{proof}[Proof Sketch for \Cref{thm:NCLB}]
We show that there exists a bipartite network $\mathsf{H}$ with less than $n$ nodes such that in each round, regardless of which nodes transmit, at most $O(\frac{1}{\log n})$ fraction of receivers receive a packet. Since at the end each receiver should receiver at least $\Omega(kB)$ bits, that is, $\Omega(k)$ packets. The $\Omega(k \log n)$ lower bound then follows. 

To prove the existence of $\mathsf{H}$, we use the probabilistic method~\cite{ProbMtd} by looking at a family of random bipartite graphs where we have $\eta=\sqrt{n}$ senders and $m=\frac{\eta \log n}{2}$ receivers. Receivers are divided into $\frac{\log n}{2}$ equal-size classes. The receivers of the $i^{th}$ class are each independently connected to exactly $2^i$ randomly chosen senders. For each fixed subset $A' \subseteq A$ of senders, let $X_{A'}$ be the random variable equal to the number of receivers that receive a packet when senders $A'$ transmit. Using simple probability calculations, we get that for any $A' \subseteq A$, $\mathbb{E}[X_{A'}]c\leq 10 \eta=O(\frac{m}{\log n})$. Most importantly, this is because, in the receiver class with degrees roughly $\frac{1}{|A'|}$, the expected number of receptions is $\Theta(\eta)$ and as we move to other classes of receivers---i.e., larger or smaller degrees---the expected number of receptions decreases exponentially. Due to the independence of edges of different receivers, we can use a Chernoff bound and infer that $Pr[X_{A'}>20\eta] \leq e^{-3\eta}$. Then, we can union bound over all $A' \subseteq A$ to get that $Pr[\nexists \, A' \subseteq A \; s.t.\; X_{A'}>20\eta] \geq 1- (2^{\eta} \cdot e^{-3\eta}) \geq 1- e^{-2\eta}$. Thus, there exists a bipartite graph $\mathsf{H}$ in this family such that transmission of no set of senders in one round can deliver packets to more than $20\eta =O(\frac{m}{\log n})$ receivers. 
\end{proof}
}
\fullOnly{
\begin{proof}
We show that there exists a bipartite network $\mathsf{H}$ with less than $n$ nodes such that in each round, regardless of which nodes transmit, at most $O(\frac{1}{\log n})$ fraction of receivers receive a packet. Then by adding a source node to this bipartite network, we complete the proof.

To prove existence of$\mathsf{H}$, we consider a distribution over a family of bipartite graphs $\mathcal{G}$ where we have $|A|=n'=\sqrt{n}$ senders and $|B|=m'=\frac{n' \log n}{2}$ receivers. The receivers are divided into $\frac{\log n}{2}$ classes, each of equal size $n'$. For each $i \in \{1, 2, \dots, \frac{\log n}{2}\}$, the receivers of class $i$ have degree exactly $2^i$ in each graph of family $\mathcal{G}$. To present the distribution, we explain how to draw a random graph from this distribution. In a random graph $G \in \mathcal{G}$, the connections are chosen randomly as follows: for each $i \in \{1, 2, \dots, \frac{\log n}{2}\}$, each receiver in the $i^{th}$ receiver class is connected to $2^i$ randomly chosen senders. The choices of different receivers are independent. Note that the size of each graph in this family is $n'(1+\frac{\log n}{2}) <n$. 

Consider a random graph $G\in \mathcal{G}$. We claim that, with probability at least $1-e^{-2n'}$, $G$ has the property that in each round at most a $O(\frac{1}{\log n})$ fraction of the receiver nodes receive a packet, regardless of which set nodes transmit. 

To prove this claim, we first study the receptions in $G$ when a fixed subset $A'$ of senders transmit. More precisely, let $X_{A'}$ be the random variable equal to number of receivers that receive a packet when senders $A'$ transmit. We first calculate $\mathbb{E}[X_{A'}]$.

Consider a receiver $v$ with degree $d$. Receiver $v$ receives a packet if and only if exactly one of its sender neighbors is in set $A'$. If $d>n'-|A'|+1$, then clearly $v$ does not receiver a packet. Suppose that $d\leq n'-|A'|+1$. Then, the probability that $G$ is such that $v$ receives a packet is exactly
{\small
\begin{equation}
\label{equ:pDelta}
\begin{split}
  P_{d}(|A'|)=\frac{\binom{|A'|}{1} \binom{n'-|A'|}{d - 1}}{\binom{n'}{d}}
  &= \frac{|A'|d}{n'}\prod_{i=1}^{d-1} (1 - \frac{|A'|-1}{n'-i}) \leq \frac{|A'|d}{n'}(1 - \frac{|A'|-1}{n'-1})^{d-1}\\
  &\leq \frac{|A'|d}{n'}\exp\left(-\frac{|A'|-1}{n'-1}(d-1)\right)\leq \frac{|A'|d}{n'}\exp\left(-\frac{|A'|-1}{n'}(d-1)\right)\\
  &= \frac{|A'|d}{n'}\exp\left(-\frac{|A'|}{n'}d\right)\exp\left(\frac{|A'|+d-1}{n'}\right)\\
  &\leq e\cdot\frac{|A'|d}{n'}\exp\left(-\frac{|A'|}{n'}d\right).
\end{split}
\end{equation}
}
For each $i \in \{1, 2, \dots, \frac{n'\log n}{2}\}$ receiver, let $X^i_{A'}$ be an indicator random variable which is each equal to $1$ iff the $i^{th}$ receiver receives a packet. We have $X_{A'}=\sum_{i=1}^{\frac{n'\log n}{2}} X^i_{A'}$. Let $d^*=2^{\lfloor\log(\frac{n'}{|A'|})\rfloor}\leq \frac{n'}{|A'|}$.
Using \Cref{equ:pDelta}, we have
{\small
\begin{equation}
\label{equ:max5}
\begin{split}
  \mathbb{E}[X_{A'}] = \sum_{i=1}^{\frac{n'\log n}{2}} X^i_{A'} = n' \sum_{i=1}^{\log n'}P_{2^i}(|A'|)
  &\leq en'\cdot\sum_{i=1}^{\log n'}\frac{|A'|2^i}{n'}
     \exp\left(-\frac{|A'|}{n'}2^i\right)\\
  &= en'\cdot\left(\sum_{i=1}^{\log d^*}\frac{|A'|2^i}{n'}
     \exp\left(-\frac{|A'|}{n'}2^i\right)+\sum_{i=\log d^*+1}^{\log n'}\frac{|A'|2^i}{n'}
     \exp\left(-\frac{|A'|}{n'}2^i\right)\right)\\
  &\leq en'\cdot\left(\sum_{j=0}^{\infty}\frac{1}{2^j}
    +\sum_{j=0}^{\infty}\frac{2^{j+1}}{e^{2^{j}}}\right)<10n'\nonumber.
\end{split}
\end{equation}
}
Note that the random variables $X^i_{A'}$ for different receivers $i$ are independent as the neighbors of different receivers are chosen independently. Thus, we can use a chernoff bound and infer that $Pr(X_{A'}>20 n')<e^{-3n'}$.
That is, when exactly nodes in set $A'$ are transmitting, with probability at least $1-e^{-3n'}$, random graph $G$ is such that at most $20n'$ receivers receive a packet. 

The total number of choices for set $A'$ is $2^{n'}$. Thus, by a union bound over all choices of set $A'$, we get that with probability at least $1-e^{-3n'}\cdot2^{n'}>1-e^{-2n'}$, the random graph $G$ is such that no set $A'$ can deliver a packet to more than $20n'$ receivers. Hence, there exists a bipartite graph $\mathsf{H}$ in this family such that no set $A'$ delivers a packet to more than $20n'$ receivers. Since there are $\frac{n' \log n}{2}$ receivers, we get that in $\mathsf{H}$, there does not exist a subset of senders which their transmission delivers a packet to more than a $\frac{40}{\log n}$ fraction of the receivers.

Now consider network $\mathsf{H}$ proven to exist. We construct radius-$2$ network $\mathsf{H}'$ from $\mathsf{H}$ by simply adding one source node $s$ and connecting $s$ to all senders. Put $k$ messages in the source node $s$. For each receiver node $u$, in order for $u$ to have all the $k$ messages, $u$ must receive at least $\Omega(kB)$ bits, i.e., $\Omega(k)$ packets. Note that this holds for any algorithm including network coding algorithms. Since receiver nodes are only connected to the sender nodes, by the choice of $\mathsf{H}$, we get that in each round at most $O(\frac{1}{\log n})$ of receivers receive a packet (any packet). Thus, it takes at least $\Omega(k\log n)$ rounds till all receivers have all the $k$ messages.    
\end{proof}
}

We remark that \Cref{thm:NCLB} can also be obtained from the proof of the $\Omega(n \log n)$ gossip lower bound proof of Gasienec and Potapov~\cite{GP02}, which itself is achieved by a reduction to the $\Omega(\log^2 n)$ lower bound of \cite{ABLP}. The proof presented here is more direct, simpler, and shorter than the proof of~\cite{ABLP}. Consequent to a preliminary writeup \cite{GHK13-throughput-arxiv} of \Cref{thm:NCLB}, Newport~\cite{Newport13} uses it to present a simpler and stronger proof of the optimal distributed single-message broadcast lower bound of \cite{KM}.

\section{Routing}\label{sec:routing}
In this section, we study the worst-case optimal routing throughput. In particular, in \Cref{subsec:schedule}, we present \emph{solitude transmission schedules} (STS), which are simple concepts that provide a more crisp and manageable way for working with routing algorithms in the bipartite networks. In \Cref{subsec:gap}, we use STSs to present a network with network coding advantage of $\Theta(\log \log n)$. 
Finally, in \Cref{subsec:routingUB}, we use STSs to present worst-case throughput optimal routing algorithms.

\subsection{Solitude Transmission Schedules}\label{subsec:schedule}
On bipartite networks, each routing-based algorithm is simply a \emph{transmission schedule}, which we formally define next. Consider a bipartite network $H=(V, E)$ with senders $A$ and receivers $B$. A (routing-based) transmission schedule $\mathcal{S}$ for $H$ is a sequence which for each round $r$, determines mutually disjoint sets $T^{m}_{r}$ for messages $m\in \{1, \dots, k\}$, where $T^m_{r}\subseteq A$ is the set of senders that transmit message $m$ in round $r$. The size of a transmission schedule $\mathcal{S}$, denoted by $|\mathcal{S}|$, is simply the number of rounds that it has. 

Even though transmission schedules are cleanly defined concepts, the fact that one needs to consider all the $k$ messages in all rounds makes the task of studying transmission schedules extremely cumbersome. To go away from this issue, we define \emph{solitude transmission schedules} which allow us to zoom in on the transmissions and the receptions of one message, while transmissions of other messages are regarded as ``noise". We show in \Cref{lem:STStrans} and \Cref{lem:stspacking} that, while STSs allow us to only work with one message (which is significantly simpler), they capture the throughput well.

\begin{mdframed}[hidealllines=true,backgroundcolor=gray!20]
\begin{definition} A \textbf{Solitude Transmission Schedule (STS)} $\mathcal{S}$ is a sequence $(i_1, A_1)$, $(i_2, A_2)$, $\ldots$, $(i_{|\mathcal{S}|}, A_{|\mathcal{S}|})$ which for each round $r$, determines a set of senders $A_r \subseteq A$ and one specific sender $i_r\in A_r$; sender $i_r$ transmits the message (of interest) and senders in $A_r \setminus \{i_r\}$ transmit ``noise". A receiver node $u$ of a bipartite network $H$ receives the message if and only if there exist a round $r$ such that the only neighbor of $u$ that is in set $A_r$ is sender node $i_r$. We say that STS $\mathcal{S}$ \emph{covers} $H$ if using $\mathcal{S}$ in $H$, all receivers receive the message. We define the \emph{weight} of an STS $\mathcal{S}$ to be $W(\mathcal{S})=\sum_{r=1}^{|\mathcal{S}|} \frac{1}{|A_r|}$. 
\end{definition}
\end{mdframed}

\medskip

First, we show that high-throughput transmission schedules lead to low-weight STSs:
\begin{lemma}\label{lem:STStrans} Let $H$ be an arbitrary bipartite networks. If there is a transmission schedule $\mathcal{S}$ for $H$ that has throughput at least $x$, then there exists an STS $\mathcal{S}'$ that covers $H$ and has $W(\mathcal{S}') \leq \frac{1}{x}$.
\end{lemma}
\begin{proof}
For each round $r$ of $\mathcal{S}$, let $T_r= \bigcup_{m \in \{1, \dots, k\}} T^m_r$. In $\mathcal{S}$, we \emph{charge} each message $m$ by $\Psi(m) = \sum_{r = 1}^{|\mathcal{S}|} |T^m_r|/|T_r|$. We have $$\sum_{ m \in \{1, \dots, k\}} \Psi(m) = \sum_{ m \in \{1, \dots, k\}} \sum_{r = 1}^{|\mathcal{S}|} |T^m_r|/|T_r| = \sum_{r = 1}^{|\mathcal{S}|}\;\;\bigg(\sum_{ m \in \{1, \dots, k\}}|T^m_r|/|T_r|\bigg) = \sum_{r = 1}^{|\mathcal{S}|} 1= |\mathcal{S}|.$$ Since $\mathcal{S}$ has throughput at least $x$, we get that $\sum_{ m \in \{1, \dots, k\}} \Psi(m)=|\mathcal{S}| \leq \frac{k}{x}$ and thus, there exists a message $m^*$ such that $\Psi(m^*) \leq \frac{1}{x}$. We transform $\mathcal{S}$ into an STS $\mathcal{S}'$ by focusing on $m^*$, using the following steps: (i) remove all the rounds in which $m^*$ is not transmitted by any sender, 
(ii) for any round $r$ such that $|T_r^{m^*}|\geq 1$, split round $r$ into $|T_r^{m^*}|$ separate rounds where in each of those $|T_r^{m^*}|$ rounds, a different node of $T_r^{m^*}$ transmits message $m^*$ while every other node in $T_r$ transmits `noise'. It is easy to see that since $\mathcal{S}$ delivers $m^*$ to every receiver, $\mathcal{S}'$ covers $H$. Also, $W(\mathcal{S}') = \Psi(m^*) \leq \frac{1}{x}$.
\end{proof}

\medskip
Next, we show (almost) the converse: Using a combinatorial packing argument, we prove that low-weight STSs, with a small additional symmetry-type requirement, can be actually turned into high-throughput transmission schedules. Later in \Cref{subsec:routingUB}, we use \Cref{lem:stspacking} to get high-throughput routing-based transmission schedules.
 
\begin{lemma}\label{lem:stspacking}
Let $H$ be an arbitrary bipartite network with $\eta$ senders. Suppose there is an STS $\mathcal{S}$ on $\eta$ senders of weight $w$ such that a random permutation of $\mathcal{S}$ covers $H$ with probability $p$. Then, there exists a routing algorithm for $H$ that achieves throughput of exactly $p/w$.
\end{lemma}
\begin{proof}
Let $\mathcal{S} = ((i_1,A_1), \ldots, (i_{|\mathcal{S}|},A_{|\mathcal{S}|}))$ be the promised STS. We design a transmission schedule $\mathcal{S}$ that tries to transmit $\eta!$ messages (one message $m_{\pi}$ for each permutation $\pi$ of the set of senders $\{1, 2, \dots, \eta\}$) in $\eta!\, w$ rounds, such that for $\eta!\, p$ messages, each of these messages is \emph{successfully} delivered to all receivers. Thus, even if the $\eta! - \eta!\, p$ \emph{unsuccessful} messages are `noise' (or empty), still $\eta!\,p$ messages are delivered successfully to all receivers, in $\eta!\, w$ rounds. For large number of messages $k$, repeating the schedule of these successful messages $\ceil{k/\eta!\,p}$ times leads asymptotically to the promised routing throughput of $p / w$. 

We first declare which senders transmit in which rounds and then make the assignment of what message is transmitted by each transmitting sender. 

For each $\ell \in [1,\eta]$, let $n_\ell$ be the number of rounds $r$ of $\mathcal{S}$ such that $|A_r|=\ell$. For each $\ell \in \{1, \dots, \eta\}$, for each set $A'$ of $\ell$ senders, we assign $n_\ell (\eta-\ell)!(\ell-1)! = n_\ell \frac{\eta!}{\ell \binom{\eta}{\ell}}$ rounds where in each of these rounds, exactly nodes of $A'$ transmit. Hence, as claimed, the total number of rounds used is 
$$\sum_\ell \binom{\eta}{\ell} n_\ell \frac{\eta!}{\ell \binom{\eta}{\ell}} = \eta! \sum_\ell \frac{n_\ell}{\ell} = \eta! \sum_r \frac{1}{|A_r|} = \eta!\, w.$$

We now determine which message is transmitted by each transmitting sender in each round using a greedy procedure. For each round $t$ of $\mathcal{S}$, let $T_t$ be the set of senders that transmit in round $t$. Also, for each round $t$ of $\mathcal{S}$ and each sender $i\in T_t$, let $M_t(i)$ be the message assigned to $i$ for transmission in round $t$. Initially, $M_t(i)=null$ for all $i \in T_t$. We iterate one by one through all permutations $\pi$ of $\{1, 2, \dots, \eta\}$ and through all rounds $r \in \{1, \dots, |\mathcal{S}|\}$ of STS $\mathcal{S}$, and each time, we search for a round $t$ in transmission schedule $\mathcal{S}$ such that the set of transmitting senders $T_t$ is exactly $\pi(A_{r})$ and for which the sender $\pi(i_{r})$ has not been assigned a message for transmission yet, i.e., $M_t(i)=null$. We then assign the sender $\pi(i_{r})$ to transmit the message $m_\pi$ at round $t$ of $\mathcal{S}$.

\begin{mdframed}[hidealllines=true,backgroundcolor=gray!20]
\centering
\begin{minipage}{15cm}
\begin{algorithm}[H]
\begin{algorithmic}[1]
\small
\Statex

\For{ each permutation $\pi$ of set $\{1, 2, \dots, \eta\}$}
	\For{$r \in \{1, \dots, |\mathcal{S}|\}$}
		\State Find a round $t$ of $\mathcal{S}$ such that $T_t= \pi(A_{r})$ and $M_t(\pi(i_{r}))=null$. \label{line:find}
		\State $M_t(\pi(i_{r}))\gets m_\pi$.
	\EndFor
\EndFor
\label{alg:STSpacking}
\end{algorithmic}
\end{algorithm} 
\end{minipage}
\end{mdframed}

\smallskip

It is clear that if the find procedure in Line~\ref{line:find} of the algorithm always succeeds, then for the produced schedule $\mathcal{S}$, the STS associated with $\mathcal{S}$ and each message $m_\pi$ is exactly $\pi(\mathcal{S}) = ((\pi(i_1),\pi(A_1)), \ldots, (\pi(i_r),\pi(A_r)))$. If $\pi(\mathcal{S})$ is one of the permutations of $\mathcal{S}$ that cover $H$, then the message $m_\pi$ is delivered to all receivers. Since there are $\eta!\, p$ such permutations, this is also the number of messages that gets delivered to all receivers. 

To complete the proof, we show that the find procedure in Line~\ref{line:find} of the algorithm always succeeds. If the greedy assignment is searching for a round $t$ such that $T_t= A'$, where $|A'|=\ell$, and where specific sender $i \in A'$ has $M(i) =null$, then the greedy assignment is processing a permutation $\pi'$ and a round $r'$ of $\mathcal{S}$ such that $\pi'(i_{r'}) = i$, $\pi'(A_{r'}) = A'$, and $|A_{r'}|=\ell$. For each of the $n_\ell$ values of $r'$ for which $|A_{r'}|=\ell$, there are exactly $1(\ell-1)!(n-\ell)!$ permutations $\pi'$ such that $\pi'(r') = i$ and $\pi'(A_{r'}) = A'$. Thus, over the course of the greedy assignment, such a $(A', i )$-find request is made exactly $n_\ell (\ell-1)!(n-\ell)!$ times. This corresponds exactly to the $n_\ell (\eta-\ell)!(\ell-1)!$ number of rounds the set $A'$ is transmitting. Hence, Line~\ref{line:find} of the algorithm will always succeed and, in fact, at the end, there will remain no $M_t(i)=null$ for any $i \in T_t$.
\end{proof}

\subsection{An $\Omega(\log \log n)$ Network Coding Gap}\label{subsec:gap}
Here, we use STSs to present a network with network coding advantage of $\Theta(\log \log n)$:

\begin{theoremR}{crl:gap}There exist bipartite networks with $n$ nodes where all receivers have degrees in range $[d/2, 2d]$, for $d>n^{0.01}$, and for which the broadcast throughput of any routing algorithm is at most $O(1/\log\log n)$ messages-per-round. For any such network, the network coding throughput is $\Theta(1)$ messages-per-round.
\end{theoremR}

We first present the key part of this theorem as \Cref{thm:RoutingLB}, which proves the existence of (almost) fixed-receiver-degree bipartite networks with optimal routing throughput of $O(1/\log\log n)$ messages-per-round. After that, with a simple comparison to network coding which achieves throughput of $\Theta(1)$ messages-per-round in such networks, we prove \Cref{crl:gap}.

\begin{theorem}\label{thm:RoutingLB} There exist a bipartite network $\mathsf{H}$ with $|A|=\eta$ senders and $|B|=\eta^{6}$ receivers, where all receiver nodes have degree in range $[\frac{9\eta}{20}, \frac{11\eta}{20}]$ and any routing-based broadcast algorithm has throughput at most $O(\frac{1}{\log \log \eta})$ messages-per-round on $\mathsf{H}$.
\end{theorem}

\begin{proof}
Using \Cref{lem:STStrans}, we get that to prove \Cref{thm:RoutingLB}, it suffices to show that there is a bipartite network $\mathsf{H}$ with the described properties such that there is no STS with weight at most $\frac{99\log \log \eta}{100}$ that covers $\mathsf{H}$. In order to do this, we consider a random distribution over a family of bipartite graphs $\mathcal{G}$ where we have $|A|=\eta$ senders and $|B|=\eta^{6}$ receivers. In each random graph $G \in \mathcal{G}$, each receiver is independently connected to each sender with probability $\frac{1}{2}$.


Standard application of Chernoff and union bounds shows that for each random graph $G \in \mathcal{G}$, with probability $1-2^{-\Omega(\eta)}$, we have that in $G$, each receiver has degree in range $[\frac{9\eta}{20}, \frac{11\eta}{20}]$. To complete the proof, we show the following: for each random graph $G \in \mathcal{G}$, with probability $1-2^{-\Omega(\eta)}$, there is no STS with weight at most $\frac{99\log \log \eta}{100}$ that covers $G$.

Since for each STS $\mathcal{S}$, we have $W(\mathcal{S}) \geq \frac{|\mathcal{S}|}{\eta}$, to prove the claim we only need to focus on STSs with at most $\eta\log \log \eta \ll \eta^2$ rounds.
Consider a fixed STS $\mathcal{S}$ such that $W(\mathcal{S}) \leq 0.99 \log \log \eta$ and the length of $\mathcal{S}$ is at most $\eta^2$. We show that the probability that this STS covers $G$ is at most $e^{-\eta^5}$. Then, we use a union bound over all such STSs to conclude the proof of the claim.

We first reorder the rounds of $\mathcal{S}$ as $(i_1, A_1), (i_2, A_2),..,(i_t, A_t)$, with $t<0.99 \eta \log \log \eta < \eta^2$, where each pair $(i_j, A_j)$ corresponds to a round in which the subset $A_j$ of A
is the set of transmitting senders, and $i_j \in A_j$ is the only sender that transmits the message. The order of these pairs is chosen greedily; $A_j$ is the one for which the cardinality of $A_j-\{i_1,i_2,..,i_{j-1}\}$ is minimized among all rounds that are still available.

Let $p$ be the minimum index such that $|A_p-\{i_1,.,,i_{p-1}\}| \geq 10 \log \eta$ (if there is no such $p$ take $p=t$). Note that the minimality in the choice of $p$ implies that
also $|A_s-\{i_1,.,i_{p-1}\}| \geq 10 \log \eta$ for all $s>p$. Moreover, let $q$ be the minimum index so that $|A_q-\{i_1,.,,i_{q-1}\}| \geq 10 \log \log \eta$ (if there is no such $q$ take $q=t$).
Note that by definition $q \leq p$. As before, the minimality in the choice of $q$ implies that $|A_s-\{i_1,.,,i_{q-1}\}| \geq 10 \log \log \eta$ for all $s>q$ (and of course
$|A_s-\{i_1,.,,i_{q-1}\}| \geq 10 \log \eta$ for all $s>p$).

%

We have $p \leq \log^2 \eta$. This is because of the following: For all $j < p$, we have $|A_j| \leq j-1+10 \log \eta$, as $|A_j-\{1_1,..,i_{j-1}\}| \leq 10 \log \eta$. Thus, if $p > \log^2 \eta$, then $W(\mathcal{S})$ would be at least $\sum_{i=1}^p 1/(i+10 \log \eta) >(1-o(1)) \log \log \eta$, which would be a contradiction.
Similarly, we have $q \leq 0.5 \log \eta$. This is because, if $q > 0.5 \log \eta$, then $W(\mathcal{S})$ would be at least $\sum_{i=1}^q 1/(i+10 \log \log \eta)=(1-o(1)) \log \log \eta$, which would again be a
contradiction.


For a fixed receiver node $u \in B$, the probability that $u$ is not connected to any of the vertices $i_1,i_2,..,i_{q-1}$ and has at least $2$ neighbors
in $A_s-\{i_1,.,,i_{q-1}\}$ for all $s \geq q$ is at least 
\begin{eqnarray}&&(\frac{1}{2})^{q} (1- p \cdot O(\log \log \eta/\log^{10} \eta)- \eta^2 \cdot O(\log \eta/\eta^{10})) \nonumber\\
\geq &&(1/\sqrt {\eta}) (1-\log^2 \eta \cdot O(\log \log \eta/\log^{10} \eta)-\eta^2 \cdot O(\log \eta/\eta^{10})) \geq 1/(2 \sqrt{\eta}).\nonumber
\end{eqnarray} 
Note that if this happens, $u$ never receives the message. Thus we get that the probability that no such receiver $u$ exists is at most $(1-1/(2\sqrt{\eta}))^{\eta^6} \ll e^{-\eta^5}$. Hence, the probability that the fixed STS $\mathcal{S}$ covers random graph $G$ is at most $e^{-\eta^5}$.

Now, the total number of possibilities for STS $\mathcal{S}$ of length at most $\eta^2$ is less than $(\eta 2^\eta)^{\eta^2} <2^{\eta^4}$. This is because, for each round $r$ of $\mathcal{S}$, there are $\eta$ options for sender $i_r$ and at most $2^\eta$ options for $A_r$. Hence, using a union bound over all such STSs, we get that the probability that there exists an STS with weight at most $\frac{99\log \log \eta}{100}$ that covers $G$ is at most $2^{\eta^4} e^{-\eta^5} < 2^{-\eta}$. Therefore, we get that the described network $\mathsf{H}$ exists. 
\end{proof}

Using \Cref{thm:RoutingLB}, we now go back to proving \Cref{crl:gap}.
\begin{proof}[Proof of \Cref{crl:gap}]
Consider the graph $\mathsf{H}$ proven to exist by \Cref{thm:RoutingLB} with $\eta=n-1$ and add a source connected to all senders. \Cref{thm:RoutingLB} implies that routing has throughput at most $O(\frac{1}{\log \log n})$. To complete the proof, we show that network coding achieves a throughput of $\Theta(1)$ in this network. Delivering messages to all senders takes just $k$ rounds. Then, divide the $k$ messages into blocks of $\Theta(\log n)$ messages. We do coding only inside each block, and deliver each block completely from the senders to receivers in $\Theta(\log n)$ rounds, thus proving the corollary. In each round (of $\Theta(\log n)$ rounds), each sender transmits a new coded packet of the block with probability $1/2$ and remains silent otherwise. It is easy to see that since receiver degrees are in range $[\frac{9\eta}{20}, \frac{11\eta}{20}]$, in each round, each receiver receives a new coded packet of the block with constant probability and thus, after $\Theta(\log n)$ rounds, w.h.p., this receiver receives $\Theta(\log n)$ coded packets and thus can decode the messages of this block. A union bound over all receivers finishes the proof.
\end{proof}

\subsection{Routing-based Broadcast Algorithms}\label{subsec:routingUB}

We now use STSs and \Cref{lem:stspacking} to obtain throughput-optimal routing algorithms.


\begin{theoremR}{lem:globalschedule}
For any bipartite network $G$ with $\eta$ senders, there exists a routing scheme with throughput $\Omega(1 / \log \eta)$ messages-per-round.
\end{theoremR}
\begin{proof} 
Consider the STS $\mathcal{S}=((1, [1,1]), (2, [1,2]), \ldots, (\eta, [1, \eta]))$. That is, the STS where in the $i^{th}$ round, the $i^{th}$ sender sends the message and senders $1$ to $i-1$ send noise. Let $H$ to be an arbitrary bipartite network with $\eta$ senders and with receivers $B \subseteq 2^{[\eta]}$ (each receiver is presented by the subset of senders to which it is connected).
We claim that for any permutation $\pi$ of senders $A=\{1, 2, \dots, \eta\}$, the STS $\pi(\mathcal{S})$ covers $H$. The theorem then directly follows from \Cref{lem:stspacking} and the observation that $W(\mathcal{S})\leq \sum_{r=1}^\eta 1/r \leq \ln \eta + 1$. 

To prove the claim, consider a receiver $r_{A'} \in B$ that is connected to the senders $A' \subseteq \{1, 2, \dots, \eta\}$. In round $t = \min_{u \in A'} \pi^{-1}(u)$ of the STS $\pi(\mathcal{S})$, node $\pi(t) \in A'$ sends and only nodes $v$ for which $\pi^{-1}(v) < \pi^{-1}(t)$ send noise. By minimality of $t$, none of these nodes $v$ are in $A'$ and thus, receiver $r_{A'}$ receives the message in round $t$. 
\end{proof}

\medskip

\noindent Note that the STS in \Cref{lem:globalschedule} is independent of network $G$. So, the resulting routing scheme works obliviously---i.e., without adapting to the topology---for any bipartite network with $\eta$ senders.

Generalizing \Cref{lem:globalschedule}, we get \Cref{lem:degschedule2,lem:degschedule3,rmrk:degschedule23} for bipartite networks with a limited range of receiver degrees, as considered in \Cref{subsec:gap}. These show the bound of \Cref{crl:gap} to be the best possible for such graphs. \shortOnly{Proofs are in the spirit of that of \Cref{lem:globalschedule} with more details, and are deferred to \Cref{app:routing}.} 

\begin{theoremR}{lem:degschedule2}
For any bipartite network $G$ with $\eta$ senders in which the receivers have degrees in $[\delta,\Delta]$ there exists a routing scheme with throughput $\Theta({1}/{(\log \frac{\Delta}{\delta} + \log \log \eta)})$.
\end{theoremR}
\fullOnly{
\begin{proof}
Let $S_1, \ldots, S_{f}$ be $f = 16 \ln \eta$ sets be independently and uniformly random chosen sets of senders $A=\{1, 2, \dots, \eta\}$, each of size $\frac{\eta}{\Delta}$. Now we consider the STS $\mathcal{S}$ that is made of two parts, i.e., $\mathcal{S}=(s_1, s_2)$, where 
\begin{itemize}
\item the first part $s_1 = ((1, [1, 1] \cup S_1), \ldots, (1, [1, 1]\cup S_f), \ (2, [1, 2] \cup S_1),\ldots,(1, [1, 2]\cup S_f),$ $\ldots,$ $\ (\frac{\eta}{\Delta \ln \eta}, [1, \frac{\eta}{\Delta \ln \eta}] \cup S_1), \ldots, (1, [1, 1]\cup S_f))$, and,
\item the second part $s_2 = ((\frac{\eta}{\Delta \ln \eta} + 1, [1, \frac{\eta}{\Delta \ln \eta}+1]), (\frac{\eta}{\Delta \ln \eta} + 2, [1,\frac{\eta}{\Delta \ln \eta}+2]), \ldots , (\frac{2 \eta \ln \eta}{\delta},[1,\frac{2 \eta \ln \eta}{\delta}]))$. 
\end{itemize}

\medskip
\noindent The weight of the first part of the schedule is $\frac{\Delta}{\eta} \cdot f \cdot \frac{\eta}{\Delta \log \eta} = O(1)$ and the second part has weight $\sum_{i = \frac{\eta}{\Delta \ln \eta}}^{\frac{2\eta \ln \eta}{\delta}} 1/i \leq \ln \frac{2 \Delta \ln^2 \eta}{\delta} + 1 = O(\log \frac{\Delta \log \eta}{\delta})$. Hence, the total weight is $W(\mathcal{S}) = O(\log \frac{\Delta \log \eta}{\delta})$.

\bigskip
Next we show that for a random permutation $\pi$, $\pi(\mathcal{S})$ covers the network $G$ with probability at least $1/2$. We first remark that for each receiver $r_{A'} \in G$, which is connected to a subset of senders $A'$, and any $i \in [1,f]$, there is an independent chance of 

$$\frac{\binom{\eta - \Delta}{\frac{\eta}{\Delta}}}{\binom{\eta}{\frac{\eta}{\Delta}}} > \left(\frac{\eta - \Delta - \frac{\eta}{\Delta}}{\eta - \frac{\eta}{\Delta}}\right)^{\frac{\eta}{\Delta}} = \left(1 - \frac{\Delta}{\eta(1-\frac{1}{\Delta})}\right)^{\frac{\eta}{\Delta}} > e^{ -(1-\frac{1}{\Delta})^{-1} } \geq e^{-2} > 1/8$$
that $\pi(S_i) \cap A' = \emptyset$. The probability that for every receiver $r_{A'}$ there is an $f_{r_{A'}} \in [1,f]$ with $\pi(S_{f_{r_{A'}}}) \cap A' = \emptyset$ is therefore at least $1 - \eta \cdot (1 - 1/8)^f > 1/4$. Furthermore for a random permutation $\pi$ there is a chance of at least 

$$1 - \frac{\binom{\eta - \delta}{\frac{\eta 2\ln \eta}{\delta}}}{\binom{\eta}{\frac{2\eta \ln \eta}{\delta}}} > 1 - (\frac{\eta - \delta}{\eta})^{\frac{2\eta \ln \eta}{\delta}} = 1 - (1 - \frac{\delta}{\eta})^{2 \frac{\eta}{\delta} \ln \eta} > 1 - \eta/4$$
that a specific receiver $r \in G$ is connected to a sender in $\pi{\frac{2 \eta \log \eta}{\delta}}$ and therefore with probability $1/4$ this holds for all receivers. 

Using a union bound we get that with probability $1/2$ both properties hold and we show next that in this case the STS $\pi(\mathcal{S})$ covers $G$. To see this, consider the receiver $r_{A'}$ that is connected to the senders in $A'$ and set $t = \min_{u \in {A'}} \pi^{-1}(u)$. If $t > \frac{n}{\Delta \log \eta}$, then $r_{A'}$ is covered by part $s_2$, namely the round in which $(t, [1, t])$ was scheduled: the node $\pi(t)$ that is sending is in $A'$ by definition of $t$. Furthermore, the nodes sending noise are senders $v$ for which $\pi^{-1}(v) < \pi^{-1}(t)$ which by minimality of $t$ are not in $A'$. Similarly, if $t \leq \frac{n}{\Delta \log \eta}$, then $r_{A'}$ is covered by part $s_1$, namely the round in which $(t,[1,t] \cup S_{f_{r_{A'}}})$ was scheduled: the node $\pi(t)$ that is sending is in $A'$ by definition. Furthermore, the nodes sending noise are senders $v$ for which $\pi^{-1}(v) < \pi^{-1}(t)$ which by minimality of $t$ are not in $A'$ and nodes in $\pi(S_{f_{r_{A'}}})$ which are also not in $A'$ by definition of $f_{r_{A'}}$. 

This shows that a random permutation of $\mathcal{S}$ covers $G$ with probability at least $1/2$ while having a weight of $\Theta(\log \frac{\Delta \log \eta}{\delta})$ which together with \Cref{lem:stspacking} result in a routing scheme with throughput $\Theta(1 / \log \frac{\Delta \log \eta}{\delta})=\Theta(\frac{1} {\log (\frac{\Delta}{\delta}) + \log \log \eta})$.
\end{proof}
}

%

\begin{theorem}\label{lem:degschedule3}
For any bipartite network $G$ with $\eta$ senders in which the maximum receiver degree is $\Delta$ there exists a routing scheme with throughput $\Theta(1 / \log \Delta)$.
\end{theorem}
\fullOnly{
\begin{proof}[Proof of \Cref{lem:degschedule3}]
The approach is almost identical to the proof of \Cref{lem:degschedule2} except that we set $f = \Delta + 1$ and change the construction of the $S_i$ sets and the switching point between $s_1$ and $s_2$. In particular, we choose the sets $S_1,S_2, \ldots, S_{\Delta+1}$ to be a partition of the senders into $f = \Delta + 1$ sets each of size at least $\frac{1}{3\Delta}$. Since for each receiver $r_{A'}$ that is connected to senders $A'$ where $|A'|\leq \Delta$, it is the case that for every permutation $\pi$ there exists an $f_r \in [1,\Delta+1]$ such that $\pi(S_{f_r}) \cap A' = \emptyset$ just as in the proof of \Cref{lem:degschedule2}. Letting $s_2$ progress up to $\eta$ furthermore guarantees that every permutation of the new STS covers $G$. 
Lastly, choosing the switching point between $s_1$ and $s_2$ to be $\frac{\eta}{\Delta^2}$ leads to a potential of $\frac{3\Delta}{\eta} \cdot f \cdot \frac{\eta}{\Delta^2} = O(1)$ for the first part and $\sum_{i = \frac{\eta}{\Delta^2}}^{\eta} 1/i \leq \ln \Delta^2 + 1 = O(\log \Delta)$ for the second part.
\end{proof}
}

\begin{remark}\label{rmrk:degschedule23}Combining \Cref{lem:degschedule2,lem:degschedule3} leads to a routing throughput of $\Theta(\frac{1}{\log \Delta \,\cdot\, \min\{\frac{\log \eta}{\delta},1\}})$.
\end{remark}

\section{Open Questions}

The results of this paper raise a number of interesting questions: Note that the routing algorithm with throughput of $\Theta(1/\log n)$ messages-per-round presented in \Cref{sec:routing} works for large number of messages. It is open whether such a throughput can be achieved for smaller number of messages. In particular, we suspect that if $k$ is at most (poly-)logarithmic in $n$, then there might be an $\Omega(k\log^2 n)$ round lower bound for routing based $k$-message broadcast algorithms. This would in essence imply a ``network coding gap'' of $\Theta(\log n)$ for this small number of messages. If true, it would be interesting to know how far this ``gap for small $k$'' extends and how it depends on the number of messages $k$. The case of $k=n$ corresponds to the well-studied gossiping problem for which the optimal routing algorithm remains open. 
\bibliographystyle{acm}
\bibliography{Bdata}
\appendix
\fullOnly{\section{Appendix}}
\shortOnly{\section{Missing Details of \Cref{sec:setup}}\label{app:setup}}
\fullOnly{\subsection{Missing Details of \Cref{sec:setup}}\label{app:setup}}

Here we explain why the limit in the definition of Broadcast Throughput (see \Cref{subsec:problem}) is well-defined.

\begin{proposition}\label{prop:limit} The limits $\lim_{k\rightarrow \infty}\frac{k}{T^R_k(G, s)}$ and $\lim_{k\rightarrow \infty}\frac{k}{T^{NC}_k(G, s)}$ defined in \Cref{subsec:problem} exist.
\end{proposition}
\begin{proof}
 First note that as each node can receive at most one message per round, $\limsup_{k\rightarrow \infty}\frac{k}{T^R_k(G, s)}$ and $\limsup_{k\rightarrow \infty}\frac{k}{T^{NC}_k(G, s)}$ are well-defined and are at most $1$. Now we argue that $\limsup_{k\rightarrow \infty}\frac{k}{T^R_k(G, s)} =\lim_{k\rightarrow \infty}\frac{k}{T^R_k(G, s)}$. A similar argument works for network coding throughput. Let $x= \limsup_{k\rightarrow \infty}\frac{k}{T^R_k(G, s)}$. Note that $x\leq 1$. For any $\eps >0$, we know that there exists a $k$ such that $|\frac{k}{T^R_k(G, s)} - x| \leq \eps/2$. We argue that for any number of messages $k' \geq k\cdot \frac{2}{\eps}$, we have $|\frac{k'}{T^R_{k'}(G, s)} - x| \leq \eps$. Divide the $k'$ messages into $\ceil{\frac{k'}{k}}$ blocks, where each block has at most $k$ messages. We can route each block in at most $k/(x-\eps/2)$ rounds. Thus, the total throughput is at least $$\frac{k'}{\ceil{\frac{k'}{k}} \cdot k/(x-\eps/2)}\geq \frac{k'(x-\eps/2)}{k'+k} \geq (x-\frac{\eps}{2})(1-\frac{k}{k'})\geq (x-\frac{\eps}{2})(1-\frac{\eps}{2})\geq x-\eps.$$
which completes the proof of existence of the limit.
\end{proof}

\shortOnly{
We now restate \Cref{prop:goingToBipartite} (claimed in \Cref{subsec:bipartite}) and prove it.

\begin{theoremR}{prop:goingToBipartite}For both routing and network coding, the worst-case optimal throughput on $n$-node bipartite networks is, up a constant factor, equal to the worst-case optimal throughput general graphs with $n+1$ nodes.
\end{theoremR}

\begin{proof}[Proof of \Cref{prop:goingToBipartite}]
To prove the theorem, we show two things: 
\begin{itemize}
\item[(a)] If there is a bipartite network with $n$ nodes for which any algorithm has bipartite broadcast throughput at most $x$ messages-per-round, then there is a network with $n+1$ nodes for which any algorithm has throughput at most $x$ messages-per-round. 

\item[(b)] If for any bipartite network with $n$ nodes we have a broadcast algorithm with throughput of at least $x$ messages-per-round, then for any general network with $n+1$ nodes we have a broadcast algorithm with broadcast throughput at least $\Theta(x)$ messages-per-round.
\end{itemize}

For part (a), simply add a source node $s$---which initially contains all $k$ messages---to the bipartite network and connect it to all senders. If a broadcast algorithm delivers $k$ messages to all nodes in at most $\frac{k}{x}$ rounds, then repeating the same transmissions in the bipartite network part gives a bipartite broadcast algorithm with throughput at least $x$.

For part (b), note that for each general graph $G$, the broadcast problem on $G$ can be broken into broadcast problems on a set of bipartite networks. In particular, if we have a broadcast algorithm with throughput $y$ for all bipartite networks with at most $n$ nodes, then we can achieve a throughput of at least $\Theta(y)$ in general graphs with $n+1$ nodes. Consider an arbitrary graph $G$ with source node $s$ and the Breadth First Search (BFS) layering of $G$ where the $i^{th}$ layer contains all the nodes at distance $i$ from the the source $s$. Each two consecutive BFS-layers define a bipartite network. We divide the messages into batches of $k'=\frac{k}{D}$. Delivering each batch from one layer to the next takes $O(k'/y)$ rounds. By spacing the progress of batches $3$ layers apart, we can pipeline different batches simultaneously and deliver all batches to all nodes in $O((D + \frac{k}{k'}) \cdot \frac{k'}{y}) = O(\frac{k}{y})$ rounds, thus achieving a throughput of $\Theta(y)$. 
\end{proof}
}
\shortOnly{\section{Missing Details \Cref{subsec:results}}\label{app:ncub}}
\fullOnly{\subsection{Missing Details \Cref{subsec:results}}\label{app:ncub}}
\begin{theorem}\label{thm:NCUB} For any network with at most $n$ nodes, there is a network coding broadcast algorithm with throughput $\Theta(\frac{1}{\log n})$.
\end{theorem}

\begin{proof}[Proof of \Cref{thm:NCUB}] Following the discussion in \Cref{subsec:bipartite}, we know that to prove the theorem, it is enough to present a broadcast algorithm with throughput $\Theta(\frac{1}{\log n})$ for bipartite networks. Fix a bipartite network $G$ with senders $S$ and receivers $R$. Divide $k$ messages into blocks of $\log n$ messages each. We use network coding to broadcast each block from senders to receivers in $\Theta(\log^2 n)$ rounds, thus achieving the claimed throughput. In the following, we explain the process for just once block. 

Using the Random Linear Network Coding analysis of \cite{Haeupler11}, it is easy to see that for each receiver $u \in R$ to be able to decode all messages of the block, it is enough if $u$ receives $\Theta(\log n)$ (different) randomly coded packets of this block. To achieve this, the $\Theta(\log^2 n)$ rounds are divided into $\Theta(\log n)$ phases, where in each round of the $i^{th}$ phase, each sender transmits a new randomly coded packet with probability $2^{-i}$ and remains silent otherwise.  
%
%

Consider a specific receiver $u$ and suppose that $u$ is connected to $d$ senders. Let $i^*=\ceil{\log d}$. Then, in each round of the phase $i^*$, $u$ receives a new packet with probability at least $\frac{d}{2^{i^*}} (1-\frac{1}{2^{i^*}})^{d-1} \geq \frac{1}{16} = \Theta(1)$. Hence, in $\Theta(\log n)$ rounds of phase $i^*$, $u$ receives $\Theta(\log n)$ coded packets with high probability. A union bound over all receivers shows that each receiver receives at least $\Theta(\log n)$ coded packets and thus, all receivers can decode all the messages of the block.
\end{proof}
The important point about the algorithm of \Cref{thm:NCUB} is that it uses different transmission probabilities to aim at different receiver degree ranges. That is, in the $i^{th}$ phase, transmissions happen with probability $2^{-i}$ and this aims at receivers with degree roughly $2^{i}$. Thus, also if we are given the promise that in a bipartite network all receiver degrees are in range $[\delta, \Delta]$, then just using the phases in range $[\floor{\log \delta}, \ceil{\log \Delta}]$ gives a broadcast algorithm with throughput $\Theta({1}/{\log {\frac{\Delta}{\delta}}})$.

%
%
%

\shortOnly{\input{NC-LB-Proof}}
\shortOnly{\input{RoutingAppendix}}

\end{document}